\documentclass{iaus}
\usepackage{graphicx}

\title[Stellar Evolution at Low Metallicity] 
{Stellar Evolution at Low Metallicity}

\author[R. Hirschi et~al]   
{Raphael Hirschi$^1$,
 Cristina Chiappini$^{2,3}$, Georges Meynet$^2$, Andr\'e Maeder$^2$, \& Sylvia Ekstr\"om$^2$}

\affiliation{$^1$Astrophysics group, Keele University, Lennard-Jones Lab., Keele, ST5 5BG, UK 
\\ email: {\tt r.hirschi@epsam.keele.ac.uk} \\[\affilskip]
$^2$Observatoire Astronomique de l'Universit\'e de Gen\`eve, CH-1290, Sauverny, Switzerland\\[\affilskip]
$^3$Osservatorio Astronomico di Trieste, Via G. B. Tiepolo 11, I - 34131 Trieste, Italia}

\pubyear{2008}
\volume{250}  
\pagerange{119--126}
\jname{Massive Stars as Cosmic Engines}
\editors{F. Bresolin, P.A. Crowther \& J. Puls, eds.}

\newcommand{\el}[2]{\ensuremath{\rm^{#2}\kern-0.8pt\rm#1}}

\begin{document}

\maketitle

\begin{abstract}
Massive stars played a key role in the early evolution of the Universe. They formed
with the first halos and started the re-ionisation. It is therefore very important
to understand their evolution. In this review, we first recall the effect of
metallicity (Z) on the evolution of massive stars. We then describe the strong impact of
rotation induced mixing and mass loss at very low $Z$. The strong mixing leads to a
significant production of primary \el{N}{14}, \el{C}{13} and \el{Ne}{22}. Mass loss
during the red supergiant stage allows the production of Wolf-Rayet stars, type Ib,c
supernovae and possibly gamma-ray bursts (GRBs) down to almost $Z=0$ for stars more massive than
60 $M_\odot$. Galactic chemical evolution models calculated with models of rotating
stars better reproduce the early evolution of N/O, C/O and \el{C}{12}/\el{C}{13}.
Finally, the impact of magnetic fields is discussed in the context of GRBs.

\keywords{Stars: mass loss, Population II, rotation, supernovae, Wolf-Rayet, 
Galaxy: evolution, gamma rays: bursts}
\end{abstract}

\firstsection 
\section{Introduction}
Massive stars started forming about 400 millions years after the Big Bang and ended the dark ages 
by re-ionising the Universe. They therefore played a key role in the early evolution of the Universe and
it is important to understand the properties and the evolution of the first stellar generations to
determine the feedback they had on the formation of the first cosmic structures. It is unfortunately not
possible to observe the first massive stars because they died a long time ago but their chemical signature 
can be observed in low mass halo stars (called EMP stars), which are so old and metal poor that 
the interstellar medium out of which these halo stars formed are thought to have been enriched
by one or a few massive stars.
Since the re-ionisation, massive stars have continuously injected kinetic energy 
(via various types of supernovae) and 
newly produced chemical elements (by both hydrostatic and explosive burning and s and r processes) 
into the interstellar medium of their host galaxy. They are thus important players for the
chemo-dynamical evolution of galaxies. Most massive stars leave a remnant at their death, either a 
neutron star or a black hole, which produce pulsar or X-ray binaries.

The evolution of stars is governed by three main parameters, which are the initial mass, 
metallicity ($Z$) and 
rotation rate. The evolution is also influenced by the presence of magnetic fields and of a close binary
companion. For massive stars ($M \gtrsim 10\ M_\odot$) around solar metallicity mass loss plays a crucial
role, in some cases removing more than half of the initial mass. Internal mixing, induced mainly by
convection and rotation also significantly affect the evolution of stars. In this review, after a summary of
the properties of low-$Z$ stars, we discuss the possible impact of rotation 
induced mixing and mass loss at low $Z$. We then present the implication of strong mixing and mass loss for
the nucleosynthesis and for galactic chemical evolution in the context of extremely metal poor stars. We also discuss the
effects of magnetic fields. We end with conclusions and an outlook.

\section{Properties of non-rotating low-$Z$ stars}
\begin{figure}[htb]
\begin{center}
 \includegraphics[width=0.5\textwidth]{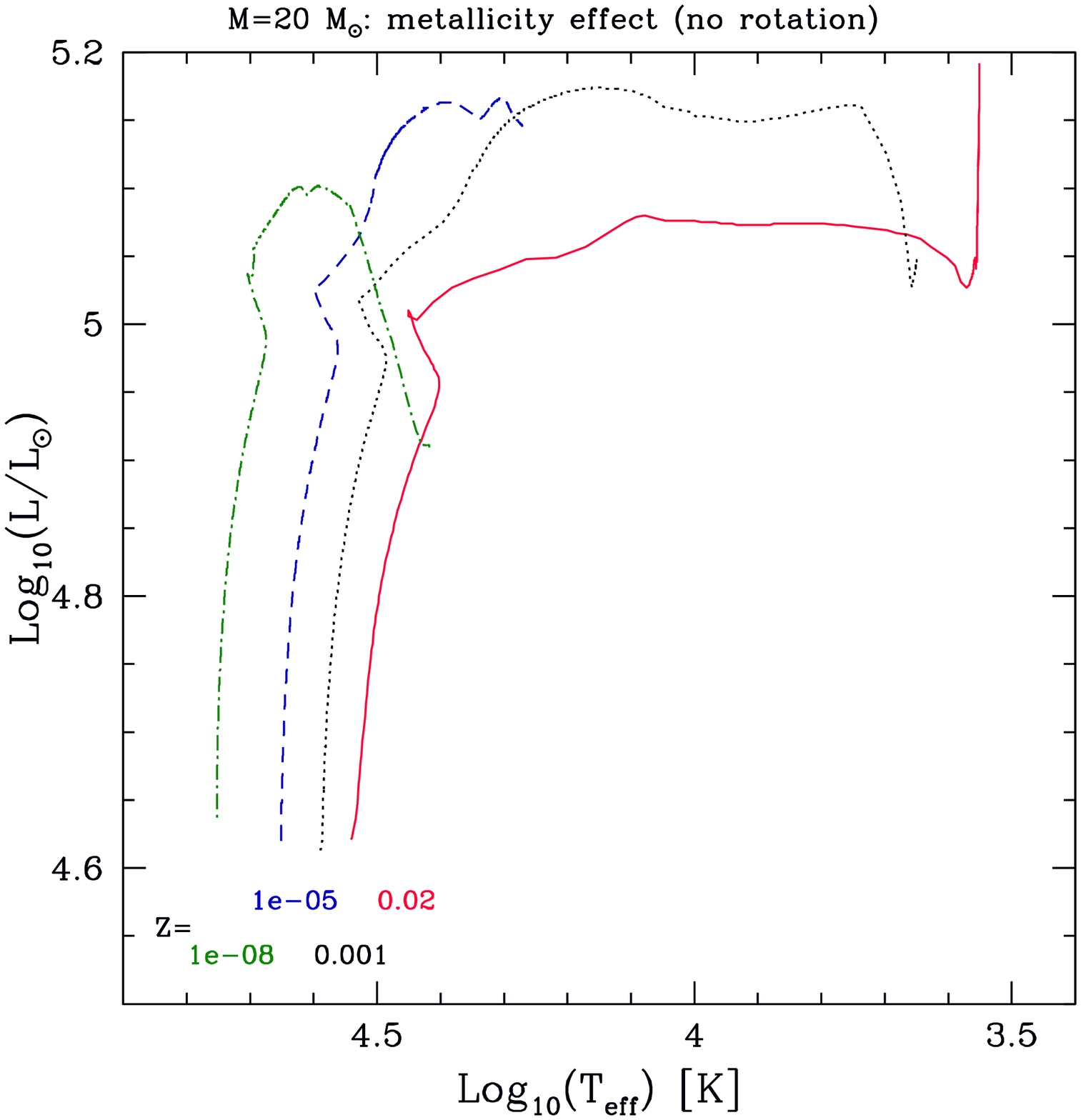}\includegraphics[width=0.5\textwidth]{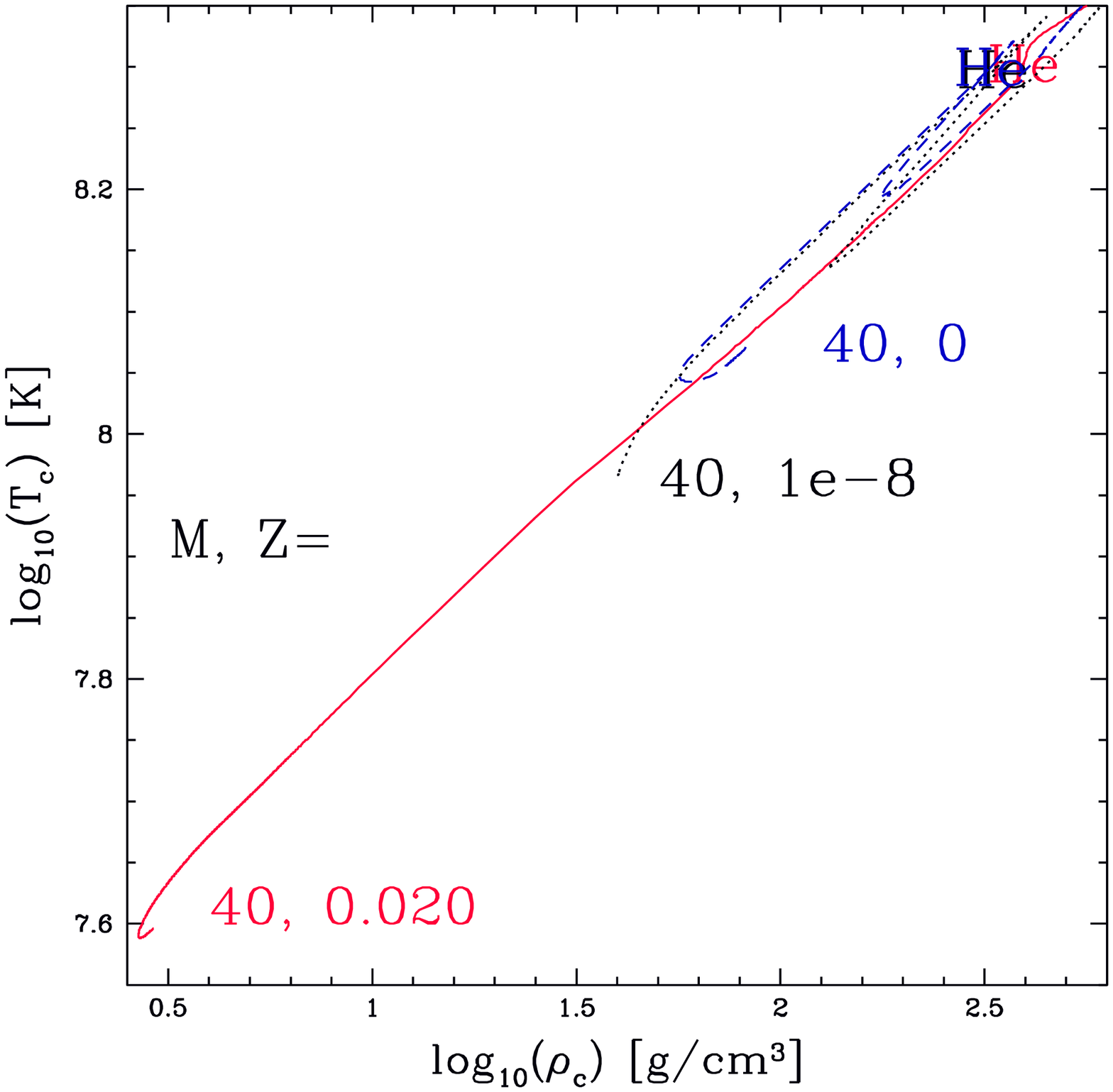} 
 \caption{{\it Left}: H-R diagram for non-rotating 20 $M_\odot$ models with $Z=10^{-8},
10^{-5}, 0.001$ and 0.02, showing that more metal poor stars have more compact envelopes 
and are less
likely to reach the red supergiant stage. {\it Right}: Central temperature versus central
density diagram for 40 $M_\odot$ models with $Z=0, 10^{-8}$ and 0.02. The evolutionary
tracks start where hydrogen burning start and the He symbols are placed at the start of
helium burning. The central conditions are much hotter and denser at very low $Z$. Note
also the different initial H-burning conditions between $Z=0$ and $10^{-8}$ 
(explained in the text).}
   \label{lowz}
\end{center}
\end{figure}
The first stellar
generations are different from solar metallicity stars due to their low metal content
or absence of it. First, low-$Z$ stars are more compact (see Fig.\,
\ref{lowz}) due to lower opacity.
Second, {\it metal free} stars burn hydrogen in a core, which is denser
and hotter due to the lack of initial CNO elements (see start of dashed curve in
Fig.\,\ref{lowz} {\it right}). 
This implies that the transition between core hydrogen and helium burning is much
shorter and smoother. Furthermore, hydrogen burns via
the pp-chain in shell burning. These differences make the metal free stars different from the
second or later generation stars (\cite{EM07}). 

Third, mass loss is metallicity dependent and therefore is expected to 
become very weak at very low metallicity. 
The metallicity ($Z$) dependence of mass loss rates is usually described using the 
formula: 
\begin{equation}
\dot{M}(Z)=\dot{M}(Z_\odot)(Z/Z_\odot)^{\alpha} \label{hirschi:mdot}
\end{equation}
The exponent $\alpha$ varies between 0.5-0.6 
(\cite{KP00ARAA,Ku02})
and
0.7-0.86 (\cite{VKL01, VdK05}) for O-type and WR stars respectively
(See \cite{MKV07} for a recent comparison between mass loss
prescriptions and observed mass loss rates). Until very recently, 
most models use at best the total metal content present at the surface of the star 
to determine the mass loss rate. However, the surface chemical composition becomes
very different from the solar mixture, due either to mass loss in the WR stage or by
internal mixing (convection and rotation) after the main sequence. It is
therefore important to know the contribution from each chemical species to opacity and
mass loss. 

Recent studies (\cite{VKL00,VdK05}) show that iron is the dominant
element concerning radiation line-driven mass loss for O-type and WR stars. 
In the case of WR stars, there is
however a plateau at low metallicity due to the contributions from light elements
like carbon, nitrogen and oxygen (CNO). 
In between the hot and cool parts of the HR-diagram, mass
loss is not well understood. Observations of the LBV stage indicate that several solar
masses per year may be lost (\cite{Sm03}) and there is no indication of a metallicity dependence.
In the red supergiant (RSG) stage, the rates generally used are still those 
of \cite{NdJ90}. More recent observations indicate that
there is a very weak dependence of dust-driven mass loss on metallicity and 
that CNO elements and especially nucleation seed components like silicon and titanium are
dominant (\cite{VL00,VL06,FG06}). 
\cite{VL05} provide recent mass loss rate prescriptions in the 
RSG stage. 
In particular, the ratio of carbon to oxygen is important to determine which
kind of molecules and dusts form. 
If the ratio of carbon to oxygen is larger than one, then carbon-rich dust would
form, and more likely drive a wind since they are more opaque than oxygen-rich
dust at low metallicity (\cite{HA07}). 

Fourth, the binary interactions are probably changed by the greater compactness of low
Z stars on the MS. Furthermore, the first generation stars below 40 $M_\odot$ do not 
evolve to the RSG stage.
\cite{MPY07} show that as Z decreases, mass transfer is more likely to take place 
after the ignition of He burning (case C) and for the very low-$Z$ stars, which do not reach the RSG stage, 
only the closest binaries would still interact.

Finally, the first stars are thought to be more massive than solar-metallicity stars 
(\cite{BL04,SOIF06}). Note that other studies suggest that both very massive and low-mass stars may form 
at $Z=0$ (\cite{NU01}). 
Since mass loss is expected to be very low at very low
metallicities, the logical deduction from these two arguments is that a large fraction of 
the first stars were
very massive at their death ($>$ 100 $M_\odot$) and therefore lead to the production of
pair-creation supernovae (PCSNe). Unfortunately,
the first massive stars died
a long time ago and will probably never be detected directly 
(see however \cite{SMWH05,TFS07}).
There are nevertheless indirect observational constraints on the first stars coming
from observations of the most metal-poor halo stars (\cite{BC05}). 
These observations do not show the peculiar chemical signature of PCSNe 
(strong odd-even effects and low zinc, see \cite{HW02}). 
This probably means that at most only a few of these very massive stars ($>$100 $M_\odot$) 
formed or that they lost a lot of mass even though their initial metal content was very 
low as discussed in the next section. Although there is no signature of PCSNe at very low
Z, they might occur in our local Universe (\cite{S06GY,LN07,WBH07}).

The topic of low Z stellar evolution is not new (see for example \cite{CC83,EE83,CBA84,A96}). 
The observations
of extremely metal-poor stars (\cite{BC05}) have however greatly increased the interest 
in very metal-poor stars. There are many recent works studying the evolution of 
metal-free (or almost) massive (\cite{HW02,LC05,UN05,MEM06}), intermediate mass
(\cite{SLL02,H04,SAMFI04,GS05}) and low mass (\cite{PCL04,W04}) stars 
in an attempt to explain the origin of the surface abundances observed.
The fate of non-rotating massive single stars at low Z is summarised in \cite{HFWLH03}
and several groups have calculated the corresponding stellar yields (\cite{HW02,CL04,TUN07}).

\section{Rotation, internal mixing and mass loss}

Massive star models including the effects of both mass loss and especially rotation better 
reproduce many observables around solar $Z$ (See contributions by Meynet and Maeder in this volume). For example, models with rotation allow chemical surface enrichments
already on the main sequence (MS), whereas without the inclusion of rotation, self-enrichments are only possible
during the RSG stage (\cite{HL002,ROTV}). Rotating star models also better reproduce the WR/O
ratio and also the ratio of type Ib+Ic to type II supernova as a function of metallicity compared
to non-rotating models, which underestimate these ratios (\cite{ROTXI}). 
The models at very low $Z$ presented here
use the same physical ingredients as the successful solar $Z$ models.
The value of 300 km\,s$^{-1}$ used as the initial rotation velocity at solar 
metallicity
corresponds to an average velocity of about 220\,km\,s$^{-1}$ on the main
sequence (MS) which is
close to the average observed value. See for instance \cite{FU82} for one
of the first surveys and the list in Meynet's contribution in this volume for the most recent surveys.
It is unfortunately not possible to observe very low $Z$ massive stars and measure their rotational 
velocity since they all died a long time ago.
Higher observed ratio of Be to B stars in
the Magellanic clouds compared to our Galaxy (\cite{MGM99}) could point
out to the fact the stars rotate faster at lower metallicities. 
Also a low-$Z$ star having the same ratio of surface velocity to critical
velocity, $\upsilon/\upsilon_{\rm crit}$ (where $\upsilon_{\rm crit}$ is the velocity for
which the centrifugal force balances the gravitational force)
as a solar-$Z$ star has a higher surface rotation velocity due to
its smaller radius (one quarter of $Z_\odot$ radius for a very low-$Z$ 20 $M_\odot$
star). In the models presented below, the initial ratio $\upsilon/\upsilon_{\rm crit}$
is the same or slightly higher than for solar Z (see \cite{H07} for more details). This
corresponds to initial surface velocities in the range of $600-800$ km\,s$^{-1}$. These
fast initial rotation velocities are supported by chemical evolution models of \cite{CH06} 
discussed in the next section.
The mass loss prescriptions used in the Geneva stellar evolution code are described in detail in
Meynet \& Maeder (2005). In particular, the mass loss rates depend 
on metallicity as $\dot{M} \sim (Z/Z_{\odot})^{0.5}$, where
$Z$ is the mass fraction of heavy elements at the surface
of the star.

How do rotation induced processes vary with metallicity? The surface layers of massive stars 
usually accelerate due to internal transport of angular momentum from the core to the envelope. 
Since at low $Z$, stellar winds are weak, this angular momentum dredged up by meridional
circulation remains in the star, and the star more easily reaches critical rotation. 
At the critical limit, matter can easily be launched into a keplerian disk which probably 
dissipates under the action of the strong radiation pressure of the star.

The efficiency of meridional
circulation (dominating the transport of angular momentum) decreases towards lower Z 
because the Gratton-\"Opik term of the vertical velocity of the outer cell is
proportional to $1/\rho$. On the other hand, shear mixing (dominating the mixing of chemical elements) 
is more efficient at low $Z$. Indeed, the star is
more compact and therefore the gradients of angular velocity are larger and the mixing 
timescale (proportional to the square of the radius) is shorter. This leads to stronger
internal mixing of chemical elements at low Z (\cite{ROTVIII}).
\begin{figure}[htb]
\begin{center}
 \includegraphics[width=0.5\textwidth]{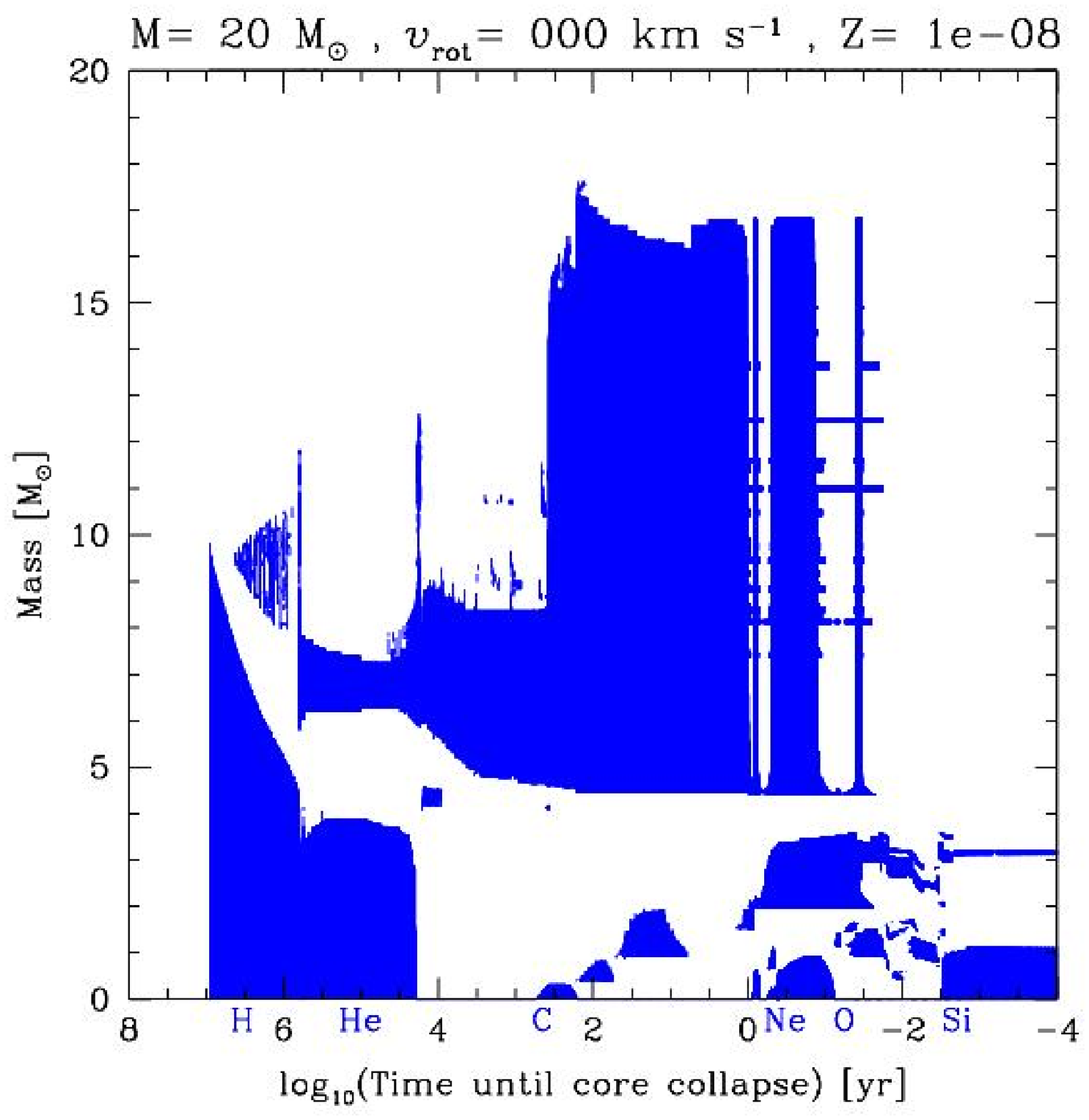}\includegraphics[width=0.5\textwidth]{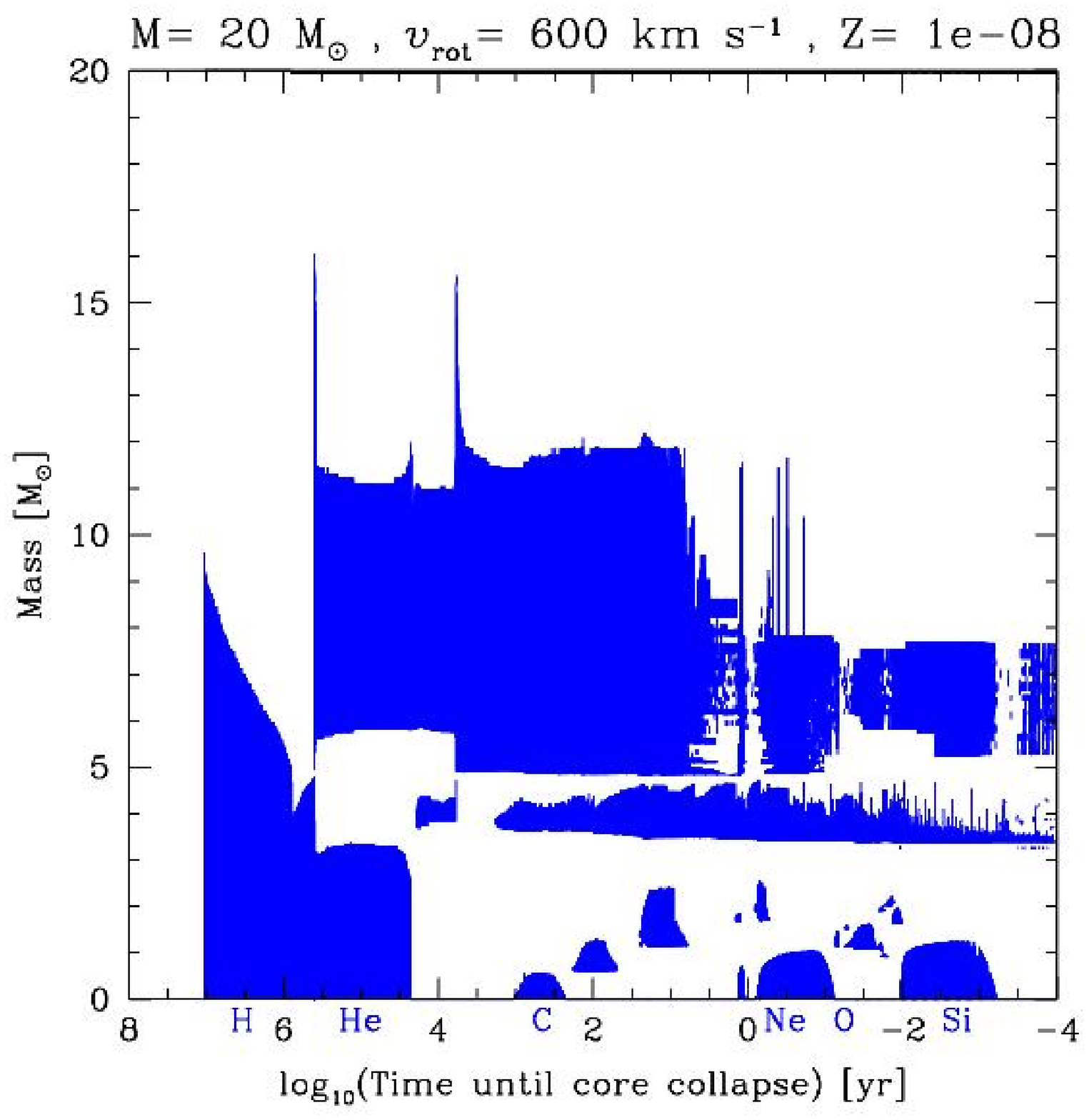} 
 \caption{Structure evolution diagram for the non-rotating ({\it left}) 
 and rotating ({\it right}) 20 $M_\odot$ models at $Z=10^{-8}$. 
Coloured areas correspond to convective zones along the Lagrangian mass 
coordinate as a function of the time left until the core collapse. 
The burning stage abbreviations are given below the time axis. Rotation
strongly affects shell H burning and core He burning.
}
\label{kip20}
\end{center}
\end{figure}

Figure \ref{kip20} shows the evolution of the convective zones in a rotating and a non-rotating 20
$M_\odot$ models at $Z=10^{-8}$. The history of convective zones (in particular the convective zones
associated with shell H burning and core He burning) is strongly affected by rotation
induced mixing. 
The most important rotation induced mixing takes place while helium is burning inside a convective core.
Primary carbon and oxygen are
mixed outside of the convective core into the H-burning shell. Once the
enrichment is strong enough, the H-burning shell is boosted (the CNO
cycle depends strongly on the carbon and oxygen mixing at such low
initial metallicities). The shell then becomes convective and leads to an important
primary nitrogen production.
In response to the shell boost, the core
expands and the convective core mass decreases.
At the end of He burning, the CO core is less massive than in 
the non-rotating model (see Fig. \ref{kip20}). 
Additional convective and rotational
mixing brings the primary CNO to the surface of the star.
This has consequences for the stellar yields.
The yield of $^{16}$O being closely
correlated with the mass of the CO core, it is therefore
reduced due to the strong mixing. 
At the same time the carbon yield is slightly increased.
The relatively "low" oxygen yields and "high" carbon
yields are produced over a large mass range at $Z=10^{-8}$ (\cite{H07}). This could
be an explanation for the possible high [C/O] ratio observed in the most
metal-poor halo stars (ratio between the surface
abundances of carbon and oxygen relative to solar; see Fig. 14 in \cite{FS6}).

\begin{figure}[ht]
\begin{center}
\includegraphics[width=0.5\textwidth]{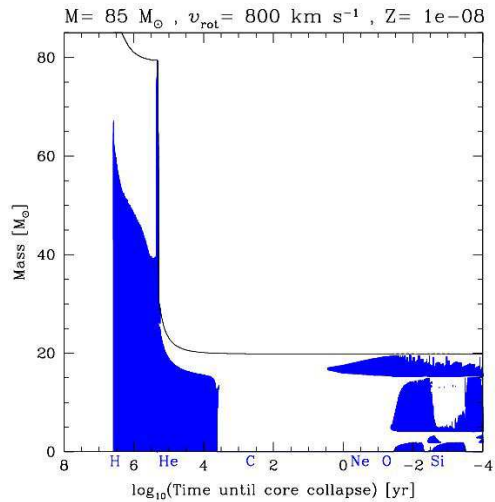}
\caption{Structure evolution diagram (same as Fig. 2) for a 85 $M_\odot$ model 
with $\upsilon_{\rm ini}=800$ km\,s$^{-1}$ 
at $Z=10^{-8}$. The top solid line shows the total mass of the
star. A strong mass loss during the RSG stage removes a large fraction of total
mass of the star.
}
\label{hirschi:kip85}
\end{center}
\end{figure}

Models of metal-free stars including the effect of rotation (\cite{EMM05})
show that stars may lose up to
10 \% of their initial mass due to the star rotating at its critical limit (also
called break-up limit). 
The mass loss due to the star reaching the critical
limit is non-negligible but not important enough to change
drastically the fate of the metal-free stars. 
The situation is very different at very low but non-zero metallicity 
(\cite{MEM06,H07}). 
The total mass of an 85\,$M_\odot$ model at $Z=10^{-8}$ is shown in Fig.
\ref{hirschi:kip85} by the top solid line. This model, like metal-free models,
loses around 5\% of its initial mass when its surface reaches break-up velocities in
the second part of the MS. At the end of core H burning,
the core contracts and the envelope expands, thus decreasing the surface
velocity and its ratio to the critical velocity. The mass loss rate becomes
very low again until the star crosses the HR diagram and reaches the RSG
stage. In the cooler part of the H-R diagram, the mass loss becomes very important. This is due to the dredge-up by the convective envelope of CNO elements to
the surface increasing its overall metallicity. The total metallicity, $Z$, is 
used in this model (including CNO elements)
for the metallicity dependence of the mass loss.
Therefore depending on how much CNO is brought up to the surface, the
mass loss becomes very large again. The CNO brought to the surface
comes from primary C and O produced in the He-burning region and from primary N produced
in the H-burning one.

Could such low-$Z$ stars undergo dust-driven winds?
%
For this to occur, the surface effective temperature needs to be low enough
(usually log(T$_{\rm eff})<3.6$) and carbon needs to be more abundant than oxygen. This last condition is fulfilled in our 85 $M_\odot$ model.
However, it is presently unclear if:
1) Extremely low-$Z$ stars reach such low effective temperatures. This
depends on the opacity and the opacity tables used in our calculations did not account for the
non-standard mixture of metals
(high CNO and low iron abundance, see \cite{Ma02} for possible effects).
2) At such low $Z$, enough metal is present to allow dust formation. Indeed, nucleation seeds (probably involving titanium) are necessary to form C-rich dust. 
There may also be
other important types of wind, like chromospheric activity-driven, pulsation-driven, 
thermally-driven or continuum-driven winds.

The fate of rotating stars at very low Z is therefore probably the following:
\begin{itemize}
\item $M < 40 \ M_{\odot}$: Mass loss is insignificant and matter is only ejected
into the ISM during the SN explosion (see contributions by
Nomoto et al, Limongi et al and Fr\"ohlich et al in this volume), which could be very energetic if
fast rotation is still present in the core at the core collapse.
\item $40\ M_\odot < M < 60\ M_\odot$: Mass loss (at critical rotation and
in the RSG stage) removes 10-20\% of the initial mass of the star. The star
probably dies as a black hole without a SN explosion and therefore the
feedback into the ISM is only due to stellar winds, which are slow.
\item $M > 60\ M_\odot$: A strong mass loss removes a significant amount of
mass and the stars enter the WR phase. These stars therefore die as type Ib/c
SNe and possibly as GRBs.
\end{itemize}

\section{Nucleosynthesis and galactic chemical evolution}
Rotation induced mixing leads to the production of primary nitrogen,
\el{C}{13} and \el{Ne}{22}. In this section, we compare the
chemical composition of our models with carbon-rich EMP stars and 
include our stellar yields in a galactic chemical evolution (GCE) model and 
compare the GCE model with observations of EMP stars.

\subsection{The most metal-poor star known to date, HE1327-2326}

\begin{figure}[ht]
\begin{center}
\includegraphics[width=0.5\textwidth]{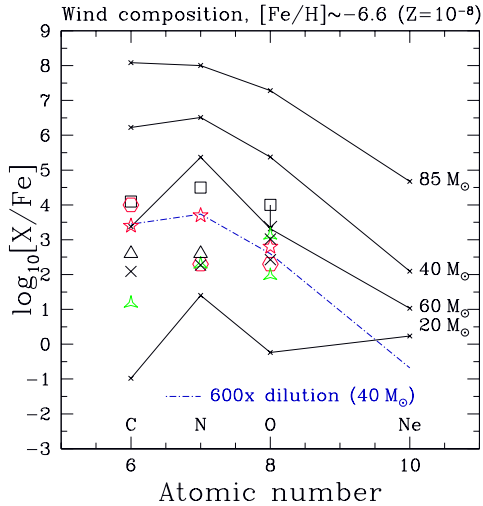}
\caption{Composition in [X/Fe] of the stellar wind for the $Z=10^{-8}$
models (solid lines).
For HE1327-2326 ({\it red stars}), the best fit for the
CNO elements is
obtained by diluting the composition of the wind of the 40 $M_\odot$
model by a factor 600 (see \cite{H07} for more details).}
\label{hirschi:cempw}
\end{center}
\end{figure}
Significant mass loss in very low-$Z$ massive stars offers an interesting
explanation for the strong enrichment in CNO elements of the most 
metal-poor stars observed in the halo of the galaxy 
(see \cite{MEM06,H07}). 
The most metal-poor star known to date, 
HE1327-2326 (\cite{Fr06}) is characterised by very high N, C and O abundances,
high Na, Mg and Al abundances, a weak s-process enrichment and depleted
lithium. The star is not evolved so has not had time to bring
self-produced CNO elements to its surface and is most likely a subgiant.
By using one or a few SNe and using a very large mass cut, 
\cite{LCB03} and \cite{IUTNM05} are
able to reproduce the abundance of most elements. 
However they are not
able to reproduce the nitrogen surface abundance of
HE1327-2326 without rotational mixing. 
The abundance pattern observed at the surface of that star
present many similarities with the abundance pattern obtained in the 
winds of very metal poor fast rotating massive star models.
HE1327-2326 may therefore have formed
from gas, which was mainly enriched by stellar winds of rotating very low
metallicity stars. In this scenario, a first generation of stars 
(PopIII) 
pollutes the interstellar medium to very low metallicities
([Fe/H]$\sim$-6). Then a PopII.5 star 
(\cite{paris05}) like the 
40 $M_\odot$ model calculated here
pollutes (mainly through its wind) the interstellar medium out of
 which HE1327-2326 forms.
This would mean that HE1327-2326 is a third generation star.
In this scenario, 
the CNO abundances are well reproduced, in particular that of
nitrogen, which according to the latest values for a subgiant (see \cite{Fr06})
is 0.9 dex higher in [X/Fe] than oxygen. 
This is shown in Fig. \ref{hirschi:cempw} where the abundances of HE1327-2326 are
represented by the red stars and the best fit is 
obtained by diluting the composition of the wind of the 40 $M_\odot$
model by a factor 600. When the SN
contribution is added, the [X/Fe] ratio is usually lower for nitrogen
than for oxygen. 
%
It is interesting to note that the very high CNO yields of the 
40 $M_\odot$ stars brings the total
metallicity $Z$ above the limit for low mass star formation
obtained in \cite{BL03}.

\subsection{Primary nitrogen and \el{C}{13}}

\begin{figure}
\begin{center}
  \includegraphics[width=0.7\textwidth]{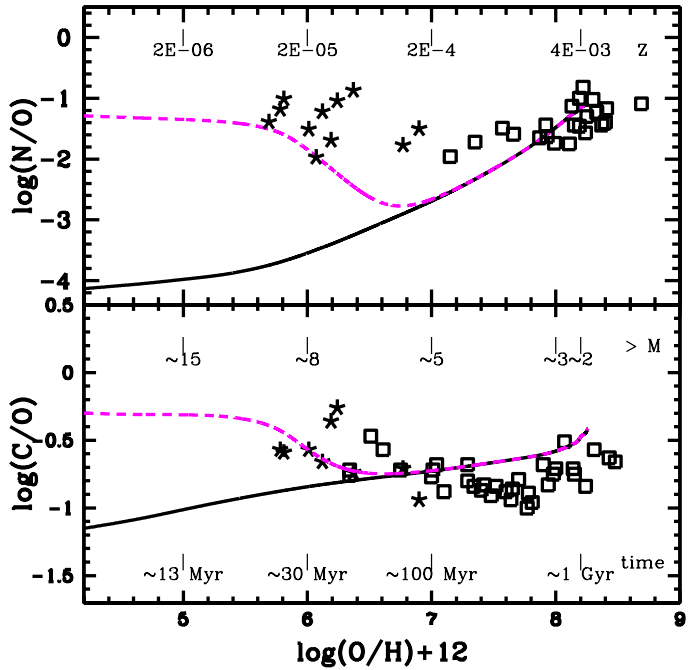}
  \caption{Chemical evolution model predictions of the N/O and C/O
evolution, in the galactic
halo, for different stellar evolution inputs. 
The solid curves show the predictions of a model without
fast rotators at low metallicities. The dashed lines show the effect 
of including a population of fast
rotators at low metallicities. 
For the data see \cite{CH06} and references therein.}
\label{CNO}
\end{center}
\end{figure}

The high N/O plateau values observed at the surface of very metal poor 
halo stars
require very efficient sources of primary nitrogen. Rotating massive stars
can inject in a short timsecale large amount of primary N at low Z. They are 
therefore very
good candidates to explain the N/O plateau observed at very low metallicity.
According to the heuristic model of \cite{CMB05}, a primary nitrogen production 
of about 0.15 $M_\odot$ per star is necessary.
Upon the inclusion of the stellar yields including the effects of fast rotation at Z$=$10$^{-8}$
in a chemical evolution model for the galactic halo with infall and outflow, 
both high N/O and C/O ratios are obtained in the very metal-poor
metallicity range in agreement with observations 
(see details in \cite{CH06b}). This model is shown in Fig. \ref{CNO}
(dashed magenta curve). In the same figure, a model computed without
fast rotators (solid black curve) is also shown. Fast rotation
enhances the nitrogen production by $\sim$3 orders of magnitude.
These results also offer a natural explanation for the large 
scatter observed in the N/O abundance ratio of {\it normal} metal-poor 
halo stars: given the strong dependency of the nitrogen yields on the 
rotational velocity of the star, we expect a
scatter in the N/O ratio which could be the consequence of the distribution 
of the stellar rotational velocities as a function of metallicity. 

As explained above, the strong production of primary nitrogen
is linked to a very active H-burning shell and therefore a smaller helium core.
As a consequence, less carbon is turned 
into oxygen, producing high C/O ratios.
Although the abundance data for C/O is still very uncertain, 
a C/O upturn at low
metallicities is suggested by observations (see \cite{A05} and references therein).
Note that this upturn is now also observed in very metal poor DLA
systems (see
the paper by M. Pettini in this volume).

\begin{figure}[!t]
\center
  \includegraphics[width=0.5\textwidth]{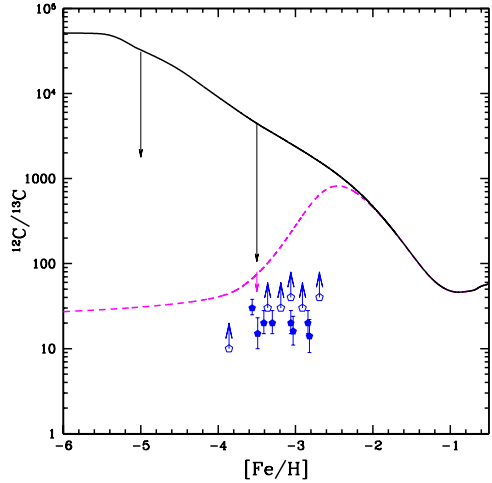}
  \caption{See a detailed description of this Fig. in Chiappini et~al. (2007)}
  \label{fig:figure2}
\end{figure}

In addition, stellar models of fast
rotators have a great impact on the evolution of the
\el{C}{12}/\el{C}{13} ratio at very low metallicities (Chiappini et~al. 2007).
In this case, we predict that, if fast rotating massive stars were
common phenomena in the early Universe, the primordial interstellar
medium of galaxies with a star formation history similar to the one
inferred for our galactic halo should have \el{C}{12}/\el{C}{13}
ratios between 30-300. Without fast rotators, the predicted
\el{C}{12}/\el{C}{13} ratios would be $\sim 4500$ at [Fe/H] $= -3.5$,
increasing to $\sim 31000$ at around [Fe/H] $= -5.0$ (see Fig.~\ref{fig:figure2}).

Current data on EMP giant normal stars in the galactic halo
(\cite{FSIX}) agree better with chemical evolution models including
fast rotators. The expected difference in the \el{C}{12}/\el{C}{13}
ratios, after accounting for the effects of the first dredge-up
(indicated by the arrows in Fig. \ref{fig:figure2}), between our predictions
with/without fast rotators is of the order of a factor of 2-3.  However, larger
differences (a factor of $\sim 60-90$) are expected for giants at [Fe/H]$=-5$ or
turnoff stars already at [Fe/H]$=-3.5$. To test our predictions, challenging
measurements of the \el{C}{12}/\el{C}{13} in more extremely metal-poor giants and
turnoff stars are required.

\subsection{Primary \el{Ne}{22} and s process at low Z}
\begin{figure}[htb]
\begin{center}
 \includegraphics[width=0.5\textwidth]{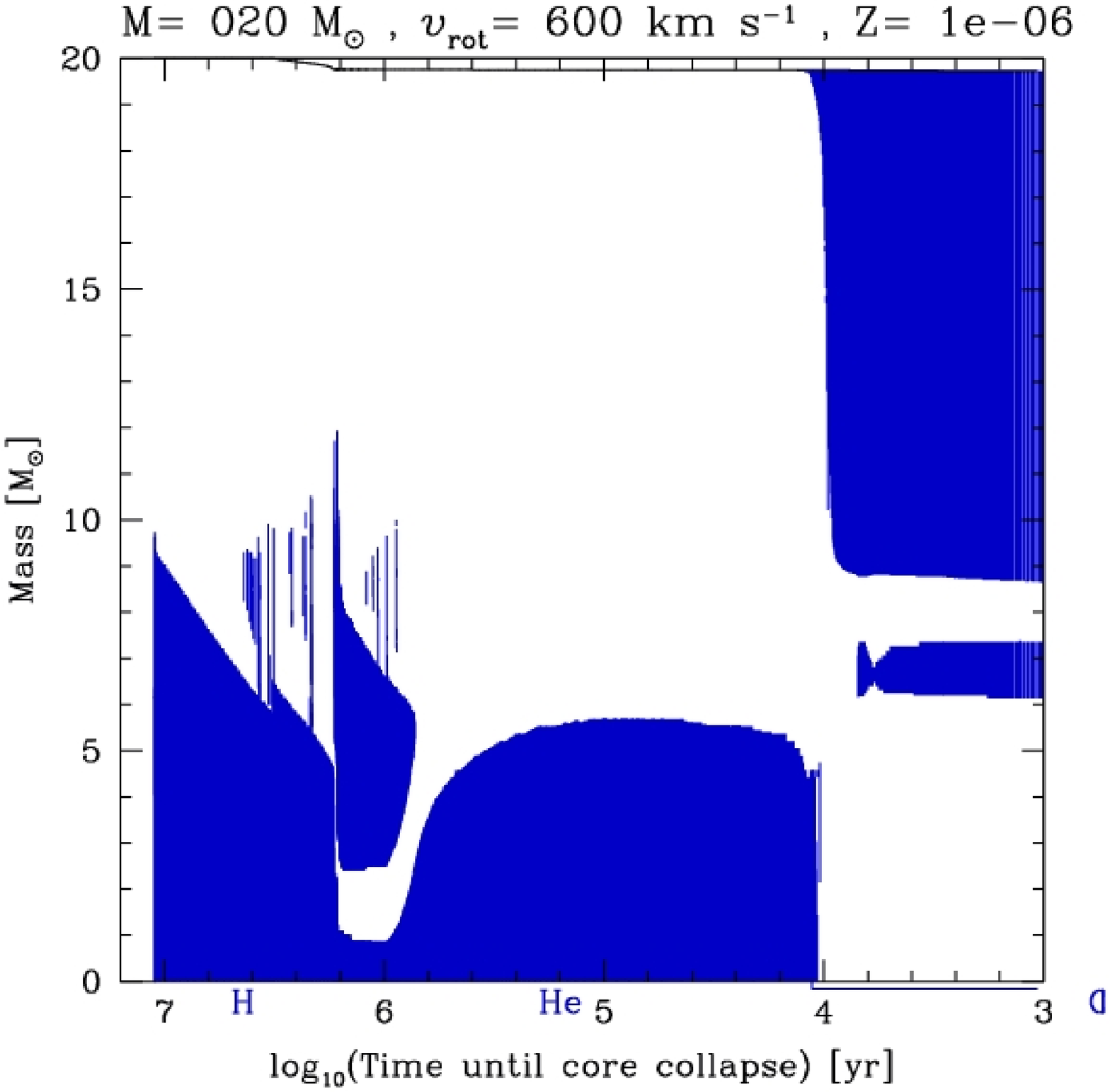}\includegraphics[width=0.5\textwidth]{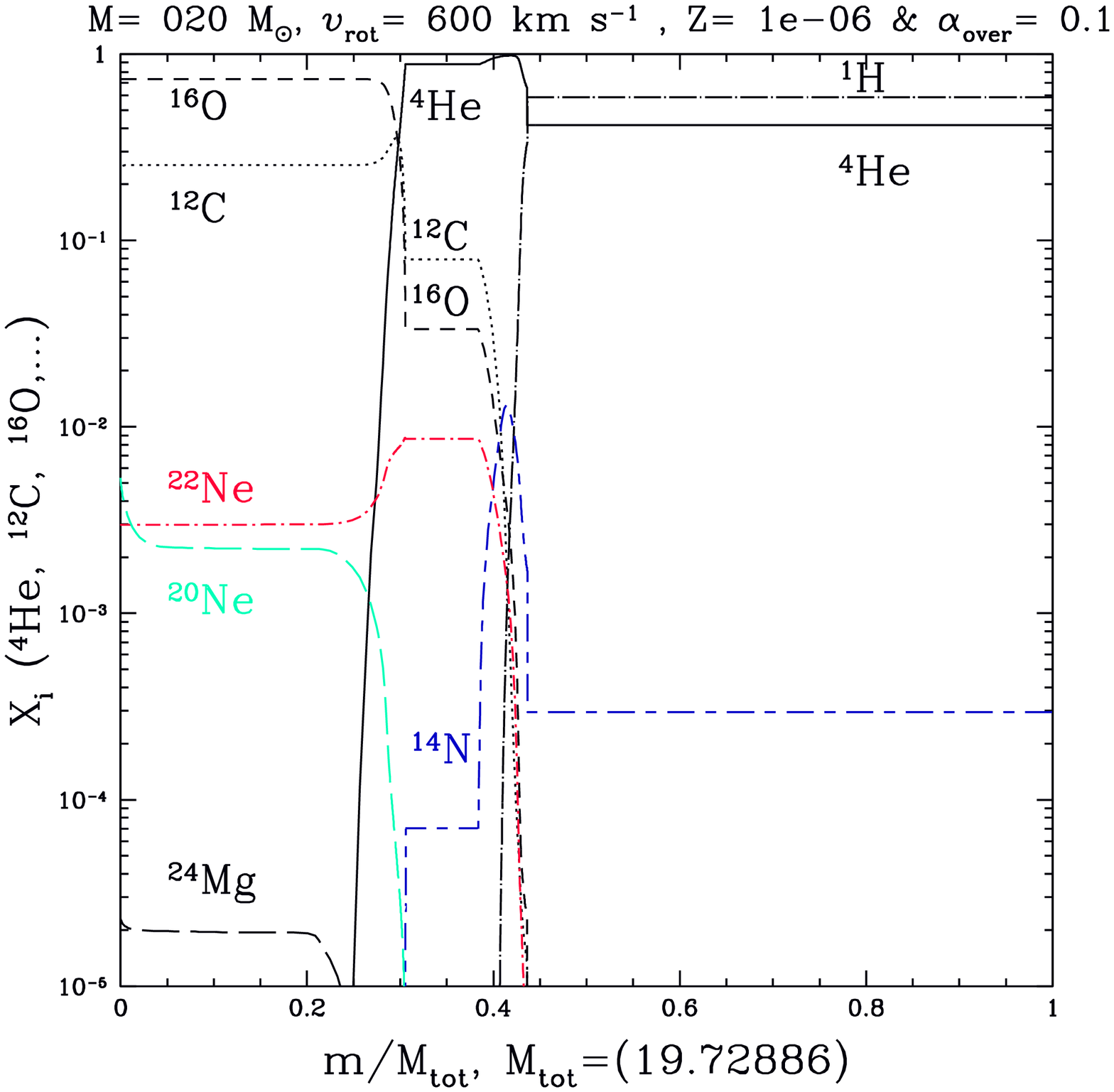} 
 \caption{{\it Left}: Structure evolution diagram (see description of Fig. \ref{kip20}) 
for a rotating 20 $M_\odot$ models 
at $Z=10^{-6}$ during H- and He-burning phases.
{\it Right}: Chemical composition at the end of core He burning. Just above the core,
one sees that the maximum abundance of \el{Ne}{22} is around 1\% in mass fraction and
at the end of core He burning, around 0.5\% is burnt in the core, providing
plenty of neutrons for s process.
}
\label{ne22}
\end{center}
\end{figure}

Models at $Z=10^{-8}$ show a production of primary \el{Ne}{22} during He burning. We also started
calculating models at different Z to determine over which Z range the primary production of \el{Ne}{22}
and also \el{N}{14} is important. In Fig. \ref{ne22}, we show the properties of a 
20 $M_\odot$ model at $Z=10^{-6}$ up to the end of He burning. Around 0.5\% (in mass
fraction) of \el{Ne}{22} is burnt during core He burning and therefore leads to a
significant neutron release. Studies are underway to determine how much s process can
be produced in these models.

\section{Magnetic fields and GRBs}
In this last section, we discuss the impact of magnetic fields. 
Models of rotating stars, which do not include the effect of magnetic fields predict
gamma-ray bursts (GRBs) at almost all Z (\cite{grb05,H07}). However, they also overestimate the
initial rotation rate of pulsars. The inclusion of the effects of magnetic fields
according to \cite{Sp02} allows a better reproduction of the initial pulsar
periods (\cite{HWS05}). Along with gravity waves, magnetic fields are also one possible 
cause for the flat rotation profile of the Sun (\cite{ROTBIV}). Although it
becomes much harder for the core to retain enough angular momentum until the core
collapse, there is still an evolutionary scenario, the so-called chemically homogeneous
evolution, leading to the production of fast
rotating cores at the pre-SN stage and therefore enabling MHD explosions and GRBs
(see Yoon et~al. 2006, \cite{WH06} and also Yoon's contribution in this volume). The theoretical GRB event rates obtained
by \cite{YLN06} are in good agreement with observations apart from the upper
metallicity limit, which is lower than the observed one (see contribution by Stanek in
this volume). \cite{YLN06} also predict that at $Z=10^{-5}$, a large fraction of
massive stars are GRB progenitors. We have calculated 40 $M_\odot$ models at
$Z=10^{-5}$ with $\upsilon_{\rm ini}/\upsilon_{\rm crit}=0.59$ and at
$Z=10^{-8}$ with $\upsilon_{\rm ini}/\upsilon_{\rm crit}=0.55$. The model at  
$Z=10^{-5}$ confirms the possibility of producing GRBs down to very low Z. However, the model at $Z=10^{-8}$
does not rotate fast enough to evolve chemically homogeneously. The difficulty of the
very low Z models to evolve chemically homogeneously is due to the weakening of the
 meridional circulation. Indeed, in models including magnetic fields, 
meridional circulation becomes the dominant term for the mixing of chemical species
(see \cite{ROTBIII}). The meridional circulation becomes weaker at low $Z$ because 
the meridional currents are less efficient in a denser medium, which is the case 
since low-$Z$ stars are more compact. 
This means that not all stars in the first stellar generations will produce GRBs in this way.
Finally, it is interesting to note that the presence of magnetic fields in metal-free stars 
may enhance mass loss 
significantly  (see contribution by Ekstr\"om et al in this volume).

\section{Conclusions and outlook}
The inclusion of
the effects of rotation changes significantly the simple picture in which 
stellar evolution at low Z is just stellar evolution without mass
loss. 
A strong mixing is
induced between the helium and hydrogen burning layers leading to a significant
production of primary \el{N}{14}, \el{C}{13} and \el{Ne}{22}. Rotating stellar models 
also predict a
strong mass loss during the RSG stage for stars more massive than 60 $M_\odot$. 
The chemical composition of the stellar winds is compatible 
with the CNO abundance observed in the most metal-poor star known to date, HE1327-2326.
GCE models including the stellar yields of these rotating star models are able to better
reproduce the early evolution of N/O, C/O and \el{C}{12}/\el{C}{13} in our galaxy. These
models predict a large neutron release during core He burning and thus a strong
possibility of an s process at very low Z. These models predict the formation of WR and
type Ib/c SNe down to almost Z=0. The inclusion of magnetic fields slows down the core
of the stars and therefore reduces the probability of producing GRBs at metallicities
around that of the Magellanic Clouds but GRBs are still predicted from single star
models down to very low Z.

Large surveys of EMP stars (SEGUE, LAMOST, and OZ surveys), of GRBs and SNe 
(Swift and GLAST satellites) and of massive stars (e.~g. VLT FLAMES survey) are underway 
and will bring more information and constraints on the evolution of massive stars at
low Z. On the theoretical side, more models are necessary to fully understand and study the complex interplay
between rotation, magnetic fields, mass loss and binary interactions at different
metallicities. Large grids of models at low Z will have many applications, 
for example to study the evolution of massive stars and their feedback in
high redshift objects like Lyman-break galaxies and damped Ly-alpha systems.



\section{Acknowledgements} R. Hirschi acknowledges financial support from EPSAM, from 
the organizers (IAU Grant), and from the Royal Society (Conference Grant round 2007/R3).


\end{document}